\documentclass{ctr}

\usepackage{ctrfont}
\usepackage{natbib}

\usepackage{url}
\usepackage{amsmath}
\usepackage{amssymb}
\usepackage[dvips]{graphicx}
\usepackage{psfrag}
\usepackage{epsfig}

\usepackage{tikz,siunitx}

\newcommand{\solidline}{\protect\tikz[baseline]{\protect\draw[solid,line width=1pt](0.0mm,0.5ex)--(6.5mm,0.5ex)}}
\newcommand{\dashedline}{\protect\tikz[baseline]{\protect\draw[dashed,line width=1pt](0.0mm,0.5ex)--(6.5mm,0.5ex)}}
\newcommand{\dotline}{\protect\tikz[baseline]{\protect\draw[dotted,line width=1pt](0.0mm,0.5ex)--(6.5mm,0.5ex)}}
\newcommand{\dotdashline}{\protect\tikz[baseline]{\protect\draw[dash dot,line width=1pt](0.0mm,0.5ex)--(6.5mm,0.5ex)}}

\title{DNS-aided explicitly filtered LES of channel flow}
\shorttitle{DNS-aided explicitly filtered LES of channel flow}
\author{H.~J.~Bae \and A.~Lozano-Dur\'an}
\shortauthor{Bae \& Lozano-Dur\'an}
\begin{document}

%%%%%%%%%%%%%%%%%%%%%%%%%%%%%%%
%!!!!!!!!!!!!!!!!!!!READ STYLE SHEET BEFORE WRITING!!!!!!!
%%%%%%%%%%%%%%%%%%%%%%%%%%%%%

%NOTE THERE IS NO ABSTRACT ALLOWED IN THESE REPORTS
\maketitle

%% SECTION NAMES:  
%% Should be lower-case except for proper nouns and abbreviations
%% Should not end with a period

\section{Motivation and objectives \label{sec:motivation}}
% Explicit filtered LES allows one-on-one correspondence with DNS with
% LES flow field.
% This allows a-posteriori distinguishablity of numerical error and
% modeling error.
% Extension method makes it possible to apply it to wall-bounded flow
% with no change in filter operator

% definition of LES and limitations of implicit filter
The equations for large-eddy simulation (LES) are formally derived by
applying a low-pass filter to the Navier--Stokes (NS) equations
\citep{Leonard1975}. However, in most numerical simulations, no
explicit filter form is specified, and the computational grid and the
low-pass characteristics of the discrete differentiation operators act
as an effective implicit filter. The resulting velocity field is then
assumed to be representative of the filtered velocity.  
Although the discrete operators have a low-pass filtering effect, the
associated filter acts only in the single spatial direction in which
the derivative is applied \citep{Lund2003}; thus each term in the
NS equations takes on a different filter form.  In addition, numerical
errors and the frequency content are uncontrolled for the implicit
filter approach, and the solutions are grid dependent
\citep{Kravchenko2000,Meyers2007}.

%#### main application (*)
% Here we have to explain the main utility of the work:
% 1) Most (filtered) DNS data is useless for aiding the construction
%    of SGS models because LES is not rigurously derived from fNS (add
%    references)
% 2) This is the case in general, but in particular it is worst for
%    wall-bounded flows where it is not clear how to filter close to
%    the wall.
% 3) By providing consistent fNS and performing explicit filter LES
%    (discretizing our fNS and adding SGS+wall model) we now can use
%    DNS data (filtered as we explain) to help improving SGS/wall
%    models.

% benefit of explicit filtered LES for SGS models
Another important limitation of implicitly filtered LES is the known
fact that the subgrid scale (SGS) tensor does not coincide with the
Reynolds stress terms resulting from filtering the NS equations due to
the implicit filter operator. This ambiguity renders direct numerical
simulation (DNS) inadequate for the development of SGS models because
of inconsistent governing equations.  On the contrary, when the
filter operator is well defined and consistent with the filtered NS
(fNS) equations, DNS data provide the necessary information to
construct exact SGS models, as demonstrated by \cite{deStefano2002}
for the simple Burger's equation.  In order to exploit the rich amount
of DNS data as a tool to devise SGS models, it is necessary to develop
an LES framework consistent with the fNS equations, that is,
explicitly filtered LES.

% past attempts at explicitly filtered LES and the benefits 
Previous works on explicitly filtered LES include the study of
\cite{Winckelmans2001}, who investigated a two-dimensional explicitly
filtered isotropic turbulence and channel flow LES to evaluate various
mixed subgrid/subfilter scale models.  \cite{Stolz2001} implemented
the three-dimensional filtering schemes of \cite{Vasilyev1998} by
using an approximate deconvolution model for the convective terms in
the LES equations. \cite{Lund2003} applied two-dimensional explicit
filters to a channel flow and evaluated the performance of explicitly
filtered versus implicitly filtered LES.  \cite{Gullbrand2003}
attempted the first grid-independent solution of the LES equations
with explicit filtering. \cite{Bose2010} further investigated the grid
independence of explicitly filtered LES with a three-dimensional
filter for turbulent channel flows.

% limitation of these past attemps for wall bounded flows
Nonetheless, the aforementioned investigations of wall-bounded
explicitly filtered LES retain some inconsistencies with the
rigorously derived incompressible fNS equations. In \cite{Lund2003}
and \cite{Gullbrand2003}, the filter operator in the wall-normal
direction was implicit even though the grid resolution was too coarse
to elude the use of a filter. In \cite{Vasilyev1998} and
\cite{Bose2010}, the filter size varied as a function of the
wall-normal distance, and the divergence-free condition for
incompressible flows was satisfied only up to a prescribed order of
accuracy. 

Recently, \cite{Bae2017} proposed a formulation of the incompressible
fNS equations that maintain consistency between the continuity
equation, the filter operator, and the boundary conditions at the wall
in the continuous limit of these equations. The method utilizes an
extension of the flow field across the wall, which allows the filter
operator to remain constant in the wall-normal direction.
\cite{Bae2017} performed a simulation of explicitly filtered
isotropic turbulence and concluded that in absense of numerical
errors, there is an one-to-one correspondence between the solutions of
the NS and fNS equations. Thus, this formulation allows a direct
comparison between DNS and explicitly filtered LES flow fields, which
in turn can inform SGS models for explicitly filtered LES. 

In this brief, we implement the formulation proposed in \cite{Bae2017}
for an explicitly filtered LES of a plane turbulent channel flow,
where the SGS model is given directly from a simultaneous simulation of
DNS starting with equivalent initial conditions. The discrepancy
due to numerical errors as well as truncation errors from
the discretization method is observed and analyzed. This analysis can
confirm whether DNS data can be used in SGS model
development for explicitly filtered LES with the consistent fNS
equations. The remainder of the brief is organized as follows. In
Section \ref{sec:framework}, the fNS equations and the extension
method for wall-bounded flows are reviewed. The numerical
experiments are introduced in Section \ref{sec:numerics}. The results
and the error analysis are given in Section \ref{sec:error}. Finally,
conclusions are offered in Section \ref{sec:summary}.

\section{Mathematical framework \label{sec:framework}}

\subsection{Filtered Navier-Stokes equations \label{sec:fil_NS}}

The incompressible NS equations and continuity condition are
\begin{equation}\label{eq:NS_eq}
{\frac{\partial {u}_i}{\partial t}} + {\frac{\partial
{u}_i{u}_j}{\partial x_j}} = - \frac{1}{\rho}{\frac{\partial {p}}{\partial
x_i}}+\nu{\frac{\partial^2 {u}_i}{\partial x_j \partial
x_j}}, \quad
{\frac{\partial{u}_i}{\partial x_i}} = 0,
\end{equation}
where the velocity components are represented by ${u}_i$, $\rho$ is the fluid density,
$\nu$ is the kinematic viscosity, and $p$ is the pressure. The filter
operator on a variable $\phi$ in integral form is defined by
\begin{equation}\label{eq:filter_in}
\bar \phi(\boldsymbol{x}) \equiv \mathcal{F}(\phi)(\boldsymbol{x})= 
\int_\Omega G(t,\boldsymbol{x},\boldsymbol{x'}) \phi(\boldsymbol{x'}) 
\mathrm{d}\boldsymbol{x'},
\end{equation}
where $\boldsymbol{x}=(x_1,x_2,x_3)$, $G$ is the filter kernel, and
$\Omega$ is the domain of integration.  When Eq. (\ref{eq:NS_eq}) is
filtered with Eq. (\ref{eq:filter_in}), the resulting equations are
\begin{equation}\label{eq:NS_eq_f}
\overline{\frac{\partial {u}_i}{\partial t}} + \overline{\frac{\partial
{u}_i{u}_j}{\partial x_j}} = - \frac{1}{\rho}\overline{\frac{\partial {p}}{\partial
x_i}}+\nu\overline{\frac{\partial^2 {u}_i}{\partial x_j \partial
x_j}}, \quad
\overline{\frac{\partial{u}_i}{\partial x_i}} = 0.
\end{equation}
The filter and differentiation operators commute when the kernel of
the filter is invariant under translation in space and time; that is,
$G(t,\boldsymbol{x},\boldsymbol{x'})=G(\boldsymbol{x}-\boldsymbol{x'})$.
When this condition is satisfied, Eq. (\ref{eq:NS_eq_f}) can be
rewritten as
\begin{equation}\label{eq:f_NS_eq}
{\frac{\partial\bar{u}_i}{\partial t}} + {\frac{\partial
\overline{{u}_i{u}_j}}{\partial x_j}} = - \frac{1}{\rho} {\frac{\partial \bar{p}}{\partial
x_i}}+\nu{\frac{\partial^2 \bar{u}_i}{\partial x_j \partial x_j}},
\quad
{\frac{\partial\bar{u}_i}{\partial x_i}} = 0,
\end{equation}
which is valid for both reversible ($\mathcal{F}^{-1}$ exists) and
irreversible filters. For reversible filters, since no information is
lost, the term $\overline{{u}_i{u}_j}$ can be expressed as a function
of $\bar{u}_i$.  For symmetric filters with Fourier transform of class
$\mathcal{C}^\infty$, the explicit form of $\overline{{u}_i{u}_j}$ as a
function of $\bar{u}_i$ has been extensively studied by
\cite{Yeo1987}, \cite{Leonard1997} and \cite{Carati2001}, among
others.  In these cases, Eq. (\ref{eq:f_NS_eq}) can be solved
independently of Eq. (\ref{eq:NS_eq}) (no closure is required), and
the solution of Eq. (\ref{eq:NS_eq}), $u_i$, is identical to the
unfiltered solution of Eq. (\ref{eq:f_NS_eq}), denoted by
$\mathcal{F}^{-1}(\bar{u}_i)$. 

When the equations are numerically integrated, another operator is
introduced, i.e., numerical discretization (e.g., the Fourier cut-off
filter for spectral discretization). An in-depth analysis of the
filter and discretization operators can be found in the work by
\cite{Carati2001}. If we denote the discretization operator by
$(\tilde{\cdot})$ and assume it commutes with the differentiation
operator, the discrete fNS equations become 
\begin{align}\label{eq:f_NS_eq_d}
&{\frac{\partial\tilde{\bar{u}}_i}{\partial t}} + {\frac{\partial
\widetilde{\overline{{u}_i{u}_j}}}{\partial x_j}} = - \frac{1}{\rho}
{\frac{\partial \tilde{\bar{p}}}{\partial
x_i}}+\nu{\frac{\partial^2 \tilde{\bar{u}}_i}{\partial x_j \partial
x_j}}.
\end{align}
The convective term of Eq. (\ref{eq:f_NS_eq_d}) is usually rearranged 
in terms of the discrete filtered velocities, and such that
\begin{align}\label{eq:f_NS_eq_d2}
&{\frac{\partial\tilde{\bar{u}}_i}{\partial t}} + {\frac{\partial
\widetilde{\tilde{\bar{u}}_i\tilde{\bar{u}}}_j}{\partial x_j}} = - \frac{1}{\rho}
{\frac{\partial \tilde{\bar{p}}}{\partial
x_i}}+\nu{\frac{\partial^2 \tilde{\bar{u}}_i}{\partial x_j \partial
x_j}}-\frac{\partial\widetilde{\mathcal{T}_{ij}}}{\partial x_j},
\end{align}
where $\mathcal{T}_{ij} = \overline{u_iu_j} -
\tilde{\bar{u}}_i\tilde{\bar{u}}_j$. 

The term $\mathcal{T}_{ij}$ can be further decomposed into the
subfilter scale (SFS) stress tensor and the SGS stress tensor, which
results from discretization errors, given that the filter and the
discretization operators commute \citep{Carati2001}.  However, for the
remainder of the brief, we consider the combined SFS and SGS stress
tensors. 

%If the discretization errors are negligible (e.g., fine grid
%resolutions/DNS), then $\tilde{u}_i \approx u_i$ and 
%$\mathcal{A}_{ij}\approx 0$. Thus, $\mathcal{A}_{ij}$ represents the
%errors due to discretization (grid resolution) and is accordingly
%named the SGS stress tensor. For the remainder of the paper, the
%discretization operator is omitted for DNS results. The term
%$\mathcal{B}_{ij}$ only depends on $\bar{\tilde{u}}_i$, and it is
%known for reversible filters. Moreover, if there is no explicit
%filter, $\mathcal{B}_{ij}=0$ and, subsequently, $\mathcal{B}_{ij}$ is
%called the subfilter scale (SFS) stress tensor. Note that only the
%term $\mathcal{A}_{ij}$ needs to be modeled since $\mathcal{B}_{ij}$
%is only a function of the discretized velocities. Another important
%remark is that, given a numerical discretization, identical
%resolutions are demanded to integrate the NS and fNS equations with
%the same degree of accuracy. The reason can be found in the $u_iu_j$
%term in $\mathcal{A}_{ij}$, which needs to be accurately computed in
%both cases and becomes the limiting factor for both equations.

One possible way of informing models for $\mathcal{T}_{ij}$ is by
using DNS data. In the most extreme case, an exact SGS model can be
produced for an explcitily filtered LES (Eq. \ref{eq:f_NS_eq_d2}) by
running a DNS (Eq. \ref{eq:NS_eq}) concurrently with the equivalent
initial condition. By assuming that the numerical errors are
negligible for the DNS velocity field, the term $\mathcal{T}_{ij}$ can
be evaluated as follows:
\begin{itemize}
\item[1.] Using the DNS velocity field, ${u_iu_j}$ is computed.
\item[2.] By filtering ${u_iu_j}$, $\overline{u_iu_j}$ is computed.
\item[3.] Using the explicitly filtered LES velocity field,
$\tilde{\bar{u}}_i\tilde{\bar{u}}_j$ is computed.

\item[4.] $\mathcal{T}_{ij}$ is calculated by $\overline{u_iu_j} -
  \tilde{\bar{u}}_i\tilde{\bar{u}}_j$.
\end{itemize}
This process allows utilization of the rich DNS data available to
inform SGS model development via examining the exact SGS term required
for a given flow field and filter and discretization operator. Note
that this would be not possible for traditional implicitly filtered
LES due to the inconsistencies in the SGS terms of the LES equations
and fNS equations. Methods such as machine learning or low-order
approximation can later be used to form SGS models. The actual
formulation of SGS models is delegated to future work. Here, we only
demonstrate that DNS data can be useful for SGS model development with
the consistent set of fNS equations. 

\subsection{Extension method \label{sec:extension}}
% problem with wall-normal filtering
The preferred form of the fNS equations for practical applications is
in Eq.  (\ref{eq:f_NS_eq}), which requires the filter operator
to commute with differentiation. This can be successfully accomplished
for unbounded flows by using the same filter operator with constant
filter width for the entire domain.  However, the presence of a wall
imposes a limitation on the support of the kernel that violates the
invariance of the filter under translation in space. If $x_2$ is the
wall-normal direction, the kernel for Eq. (\ref{eq:filter_in}) is
inevitably of the form $G(x_2,\boldsymbol{x}-\boldsymbol{x'})$ since
the integration is bounded by the presence of the wall. Therefore, the
filter and differentiation operators do not commute and Eq.
(\ref{eq:f_NS_eq}) does not apply.

\begin{figure}
\begin{center}
\vspace{0.5cm}
\includegraphics[width=0.7\textwidth]{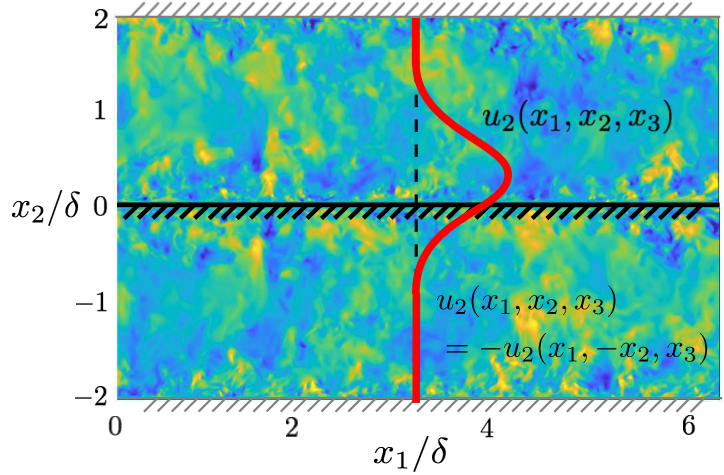}
\caption{Illustration of the extension method in Eq.
(\ref{eq:extension}) for the wall-normal velocity component of a
turbulent channel flow. Solid line depicts the kernel of the
filter operator, which can extend beyond the wall.
\label{fig:filtering_y}}
\end{center}
\end{figure}

% extended the flow
For flat walls, \cite{Bae2017} proposed to resolve this limitation by
extending the flow in the wall-normal direction, allowing for a
uniform filter in the near-wall region as illustrated in Figure
\ref{fig:filtering_y}.
Henceforth, we consider the flow over a smooth flat, wall with
$x_1,x_2,$ and $x_3$ signifying the streamwise, wall-normal, and
spanwise directions, respectively. Quantities evaluated at the wall,
located at $x_2 = 0$, are denoted by $(\cdot)|_w$. The flow is
then extended below the wall as
\begin{equation}
u_1(-x_2) = u_1(x_2),\quad u_2(-x_2) = -u_2(x_2), \quad u_3(-x_2) =
u_3(x_2), 
\label{eq:extension}
\end{equation}
where $x_1$ and $x_3$ are omitted for simplicity. In this manner,
$\partial u_i/\partial x_i = 0$ is also satisfied in the extended
domain, preserving incompressibility. Note that the velocities in the
extended domain are also consistent with the NS equations. The
extension provided by Eq. (\ref{eq:extension}) removes the limitation
on the support of the kernel previously imposed by the wall, and Eq.
(\ref{eq:f_NS_eq}) can be formally obtained for flows over flat walls.
Note that, by symmetry, the filtered velocity field also satisfies
\begin{equation}
\bar u_1(-x_2) = \bar u_1(x_2),\quad \bar u_2(-x_2) = -\bar u_2(x_2),
\quad \bar u_3(-x_2) = \bar u_3(x_2).
\label{eq:extension_ubar}
\end{equation}
Another possible extension consistent with the incompressibility
condition is 
\begin{equation}
u_1(-x_2) = -u_1(x_2),\quad u_2(-x_2) = u_2(x_2), \quad u_3(-x_2) =
-u_3(x_2).
\label{eq:extension2}
\end{equation}
However, this extension is not a compatible form, as the extended
velocities do not satisfy the NS equations.

% boundary conditionns
The boundary conditions for the filtered velocity field are derived
by consistency with the filter operator,
\begin{equation}
\label{eq:bc}
\bar{u}_{i}|_w = \mathcal{F}(u_{i})|_w.
\end{equation}
Then, the boundary conditions for the fNS equations reduce to the
usual no-slip condition for the wall-normal filtered velocity and the
filter-consistent boundary condition for the wall-parallel filtered
velocities
\begin{equation}
\begin{gathered}
\label{eq:bc_uw}
%\left.\frac{\partial\bar{u}_1}{\partial x_2}\right|_w = 0,\quad \bar{u}_2|_w =
%0,\quad \left.\frac{\partial\bar{u}_3}{\partial x_2}\right|_w = 0.
\mathcal{F}^{-1}\left.\left(\bar{u}_1\right)\right|_w = 0\text{ with }
\bar{u}_1(-x_2) = \bar{u}_1(x_2),\\
\bar{u}_2|_w = 0,\\
\mathcal{F}^{-1}\left.\left(\bar{u}_3\right)\right|_w = 0\text{ with }
\bar{u}_3(-x_2) = \bar{u}_3(x_2).
\end{gathered}
\end{equation}
This boundary condition extends to explicitly filtered LES, and the
extension method can be applied to the LES equations.

\section{Numerical experiment \label{sec:numerics}}

% Channel cases
% LES1,2,3 --> Grid resolution 1,0.5,0.25x that of DNS
% DNS feeds in the SGS

%--------------------------------------------------------%
\begin{table} 
\begin{center} 
\setlength{\tabcolsep}{12pt}
\begin{tabular}{l l c c c c c c} 
Case  & Equations   & $\langle Re_\tau\rangle_t$ & $\Delta_1^+$ & 
$\Delta_{2,\min}^+$ & $\Delta_{2,\max}^+$        & $\Delta_3^+$ \\ 
\hline
\hline
DNS   & NS          & $186$                      & 8.8          &
1.3                 & 19.4                       & 8.8          \\
LES1  & fNS         & $186$                      & 8.8          &
1.3                 & 19.4                       & 8.8          \\
LES2  & fNS         & $184$                      & 17.7         &
2.8                 & 38.4                       & 17.7         \\
LES4  & fNS         & $182$                      & 35.3         &
5.6                 & 76.8                       & 35.3         \\
\hline 
\end{tabular} 
\end{center}
\caption{Tabulated list of cases. 
\label{tab:cases}}
\end{table}
%--------------------------------------------------------%

We perform a set of plane turbulent channel DNS and explicitly
filtered LES listed in Table \ref{tab:cases} at
friction Reynolds number $Re_\tau=u_\tau\delta/\nu\approx180$, where
$u_\tau$ is the friction velocity at the wall. The simulations are
computed with a staggered second-order finite difference
\citep{Orlandi2000} and a fractional-step method \citep{Kim1985} with
a third-order Runge-Kutta time-advancing scheme \citep{Wray1990}. The
code has been validated in previous studies in turbulent channel flows
\citep{Lozano2016,Bae2018}. The size of the channel is $2\pi
\delta\times 2\delta\times \pi \delta$ in the streamwise, wall-normal,
and spanwise directions, respectively.  Periodic boundary conditions
are imposed in the streamwise and spanwise directions. The channel
flow is driven by imposing a constant mean pressure gradient. 

%--------------------------------------------------------%
\begin{figure}
\begin{center}
\vspace{0.5cm}
\includegraphics[width=1.0\textwidth]{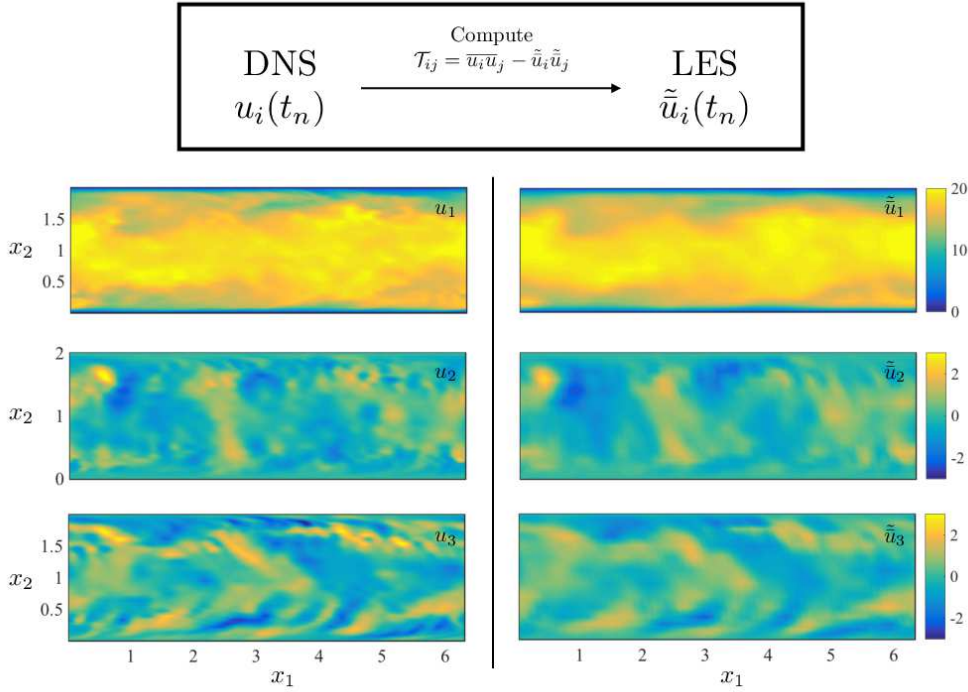}
\caption{Schematic of how explicitly filtered LES is computed from DNS
data. The term $\mathcal{T}_{ij}$ is computed at each time step from
the corresponding DNS flow field and then applied to the explicitly
filtered LES equations. Left panels show an instantaneous snapshot of
the DNS streamwise, wall-normal and spanwise velocities, and the right
panels show the corresponding instantaneous snapshot of LES1 at
$t=5\delta/u_\tau$. 
\label{fig:schematic}}
\end{center}
\end{figure}
%--------------------------------------------------------%

The DNS grid resolutions in the streamwise and spanwise directions are
uniform with $\Delta_1^+,\Delta_3^+ \approx 8.8$, where the
superscript $+$ denotes wall units given by $\nu$ and $u_\tau$.
Non-uniform meshes are used in the normal direction, with the grid
stretched toward the wall according to a hyperbolic tangent
distribution. The height of the first grid cell at the wall is
$\Delta_2^+ \approx 1.3$. This grid resolution is coarser in the
wall-normal direction compared with those in previous DNS cases \citep{Kim1987}, but for the
purpose of our study, only the comparison between the LES and DNS
cases listed here is significant. The three LES cases have grid
spacings that are 1, 2 and 4 times that of the DNS. The reversible
filter operator was chosen to be the differential filter
\citep{Germano1986}
\begin{equation}
G(\boldsymbol{x}-\boldsymbol{x}';a) = \frac{1}{4\pi
a^2}\frac{\exp(-|\boldsymbol{x}-\boldsymbol{x}'|/a)}{|\boldsymbol{x}-\boldsymbol{x}'|},
\end{equation}
with filter parameter $a = 0.1$. The parameter $a$ is related to the
filter width, $\Delta_f$, by $a = \Delta_f^2/40$ based on an
equivalent second moment with a spherical top-hat filter of radius,
$\Delta_f/2$. The filter size was chosen to be large enough to damp
the high-frequency content such that the numerical errors emanating
from inaccurate high-frequency solutions of the discrete derivative
operators are controlled \citep{Carati2003}. 

All cases were started from an equivalent initial condition, where the
initial condition of the simulations was computed by performing a
preliminary simulation in DNS resolution for at least 100 eddy
turnover times, defined as $\delta/u_\tau$. This
initial condition is then filtered and interpolated to the grid
resolution for each case. The interpolation, which serves as the
effective discretization operator, is given by a second-order
linear interpolation. Starting from the equivalent initial condition,
all cases were run for $50\delta/u_\tau$. For the LES cases, the term
$\mathcal{T}_{ij}$ is computed at each time step following the
procedure given in Section \ref{sec:fil_NS} from the DNS flow field at
the same time step (Figure \ref{fig:schematic}). The instantaneous
$u_\tau$ is denoted by $u_\tau^{\mathrm{DNS}}(t)$ and
$u_\tau^{\mathrm{LES}}(t)$ if distinction is necessary.
Time-averaged values after transients are given by
$\langle\cdot\rangle_t$.

Note that the second-order linear interpolation discretization on a
stretched mesh does not commute with the second order finite
differentiation operator; thus, additional numerical commutation
errors are expected. However, this is not an artifact of explicit
filtering but rather the result of discretization. Higher-order
interpolation methods or quadrature rules may be employed to mitigate
this commutation error, and this will be investigated in future works.

\section{Error assessment \label{sec:error}}

\begin{figure}
\begin{center}
\vspace{0.5cm}
\includegraphics[width=0.48\textwidth]{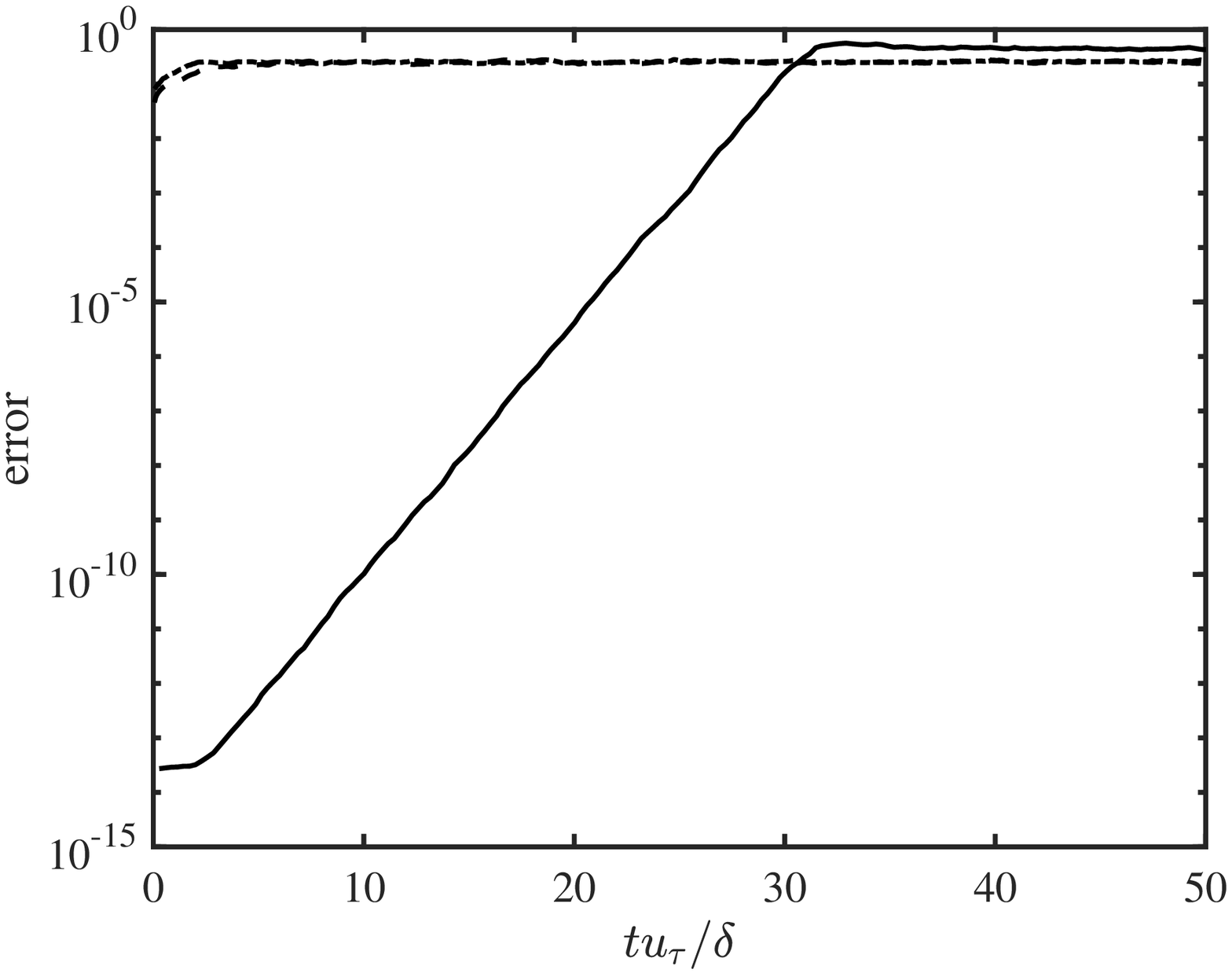}
\hspace{0.2cm}
\includegraphics[width=0.48\textwidth]{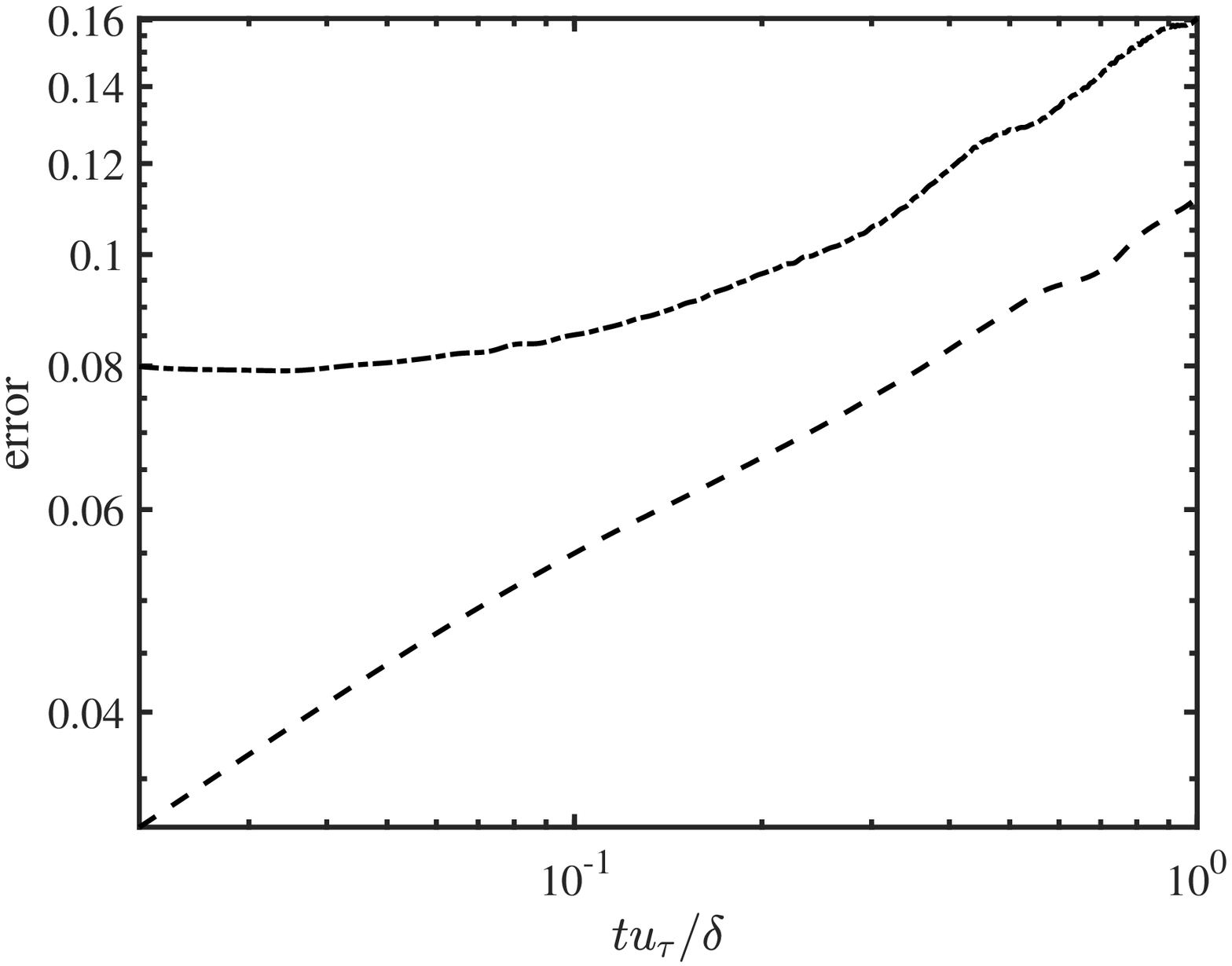}
\caption{
Error in the instantaneous velocity field as a function of time  for
LES1 (\solidline), LES2 (\dashedline), and LES4 (\dotdashline)
compared with DNS in (a) semi-log scale and (b) log-log scale (LES1
excluded).
\label{fig:error_full}}
\end{center}
\end{figure}

Owing to the chaotic nature of turbulence, the small numerical errors
that arise at each time step are expected to accumulate and affect the
solution significantly for long times. Hence, we compute the
associated error of LES as a function of time. The error for the
instantaneous streamwise velocity field can be computed as
\begin{equation}
\text{error}(t) = \frac{\left[\int_{\mathcal{V}}
\left( u_1^{\mathrm{LES}}/u_\tau^{\mathrm{LES}} 
- u_1^{\mathrm{DNS}}/u_\tau^{\mathrm{DNS}}\right)^2 
\mathrm{d}V\right]^{1/2}}
{\left[\int_{\mathcal{V}}
\left(u_1^{\mathrm{DNS}}/u_\tau^{\mathrm{DNS}}\right)^2
\mathrm{d}V\right]^{1/2}},
\end{equation}
where $V$ is the volume of the computational domain and
$u_i^{\mathrm{LES}}$ is computed by reversing the differential filter
on the instantaneous velocity field from the explicitly filtered LES.
The error of unfiltered quantities, as opposed to the
filtered ones, are examined since we are interested in predicting DNS values
rather than the filtered counterparts.  Figure \ref{fig:error_full}
shows the error as a function of time for the three LES cases.  For
case LES1, where the discretization error is negligible, the error in
the instantaneous velocity profile grows exponentially until it
saturates at $30\delta/u_\tau$ (Figure \ref{fig:error_full}(a)). The
initial plateau for $tu_\tau/\delta < 2$ occurs owing to the error being
less than machine precision. The exponential growth of error is
expected in a chaotic system with positive Lyapunov exponents, such as
turbulent channel flow \citep{Eckmann1985}.  The initial machine
precision error between the DNS and explicitly filtered LES cases
shows
that the LES equations employed in this work are consistent with the NS
equations and, following the argument in Section \ref{sec:fil_NS}, DNS
data can be useful in informing the SGS models. Cases LES2 and LES4
show a first-order growth in error that saturates around
$2\delta/u_\tau$ (Figure  \ref{fig:error_full}(b)).  This is possibly
due to the commutation error between the interpolation
(discretization) operator and the differentiation operator mentioned
in Section \ref{sec:numerics}.   
%--------------------------------------------------------%
%--------------------------------------------------------%

In the remainder of the section, we study the error in the mean
velocity profile and turbulence intensities for DNS-aided explicitly
filtered LES.

\subsection{Mean velocity profile \label{sec:mean}}

\begin{figure}
\begin{center}
\vspace{0.5cm}
\includegraphics[width=0.48\textwidth]{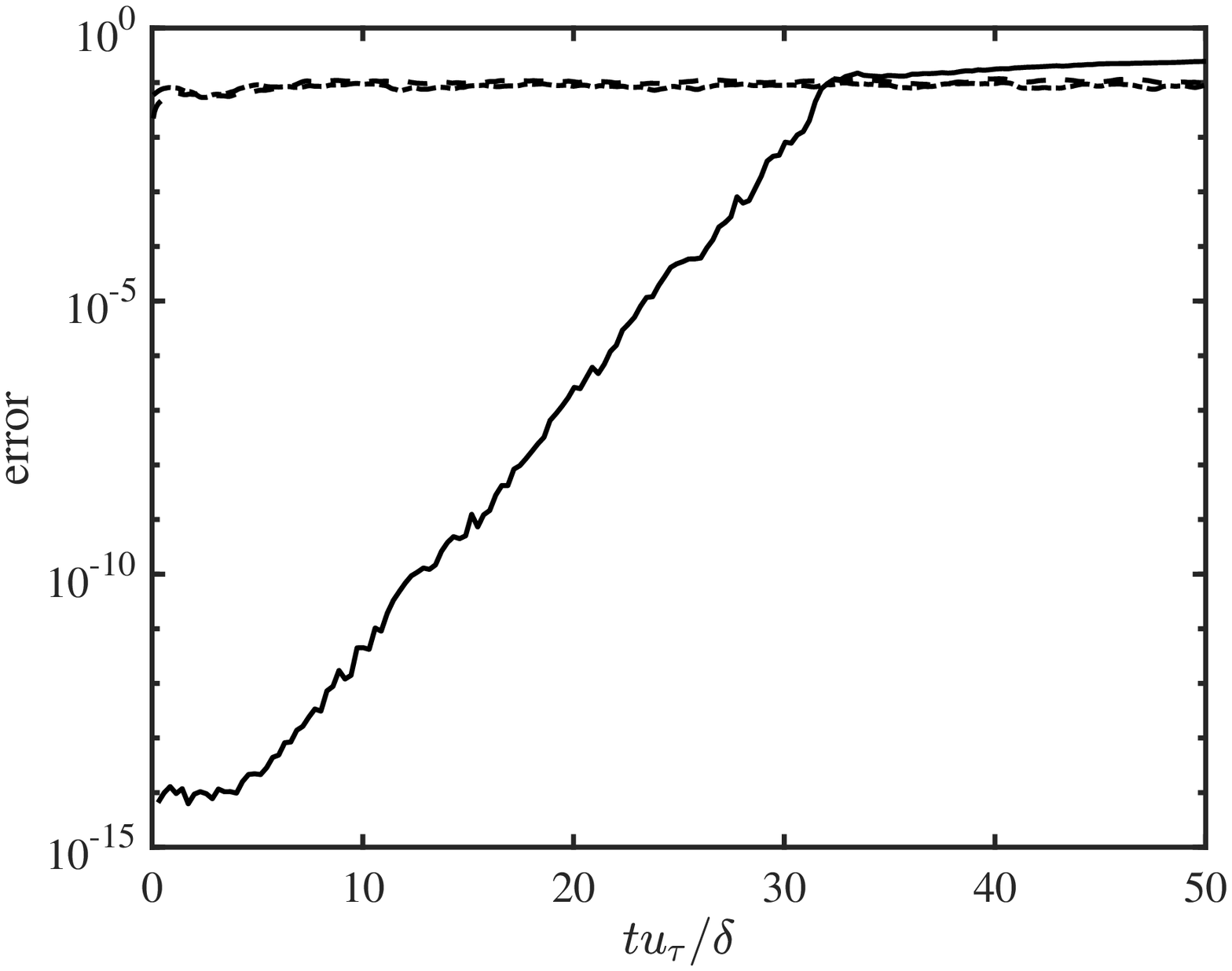}
\hspace{0.2cm}
\includegraphics[width=0.48\textwidth]{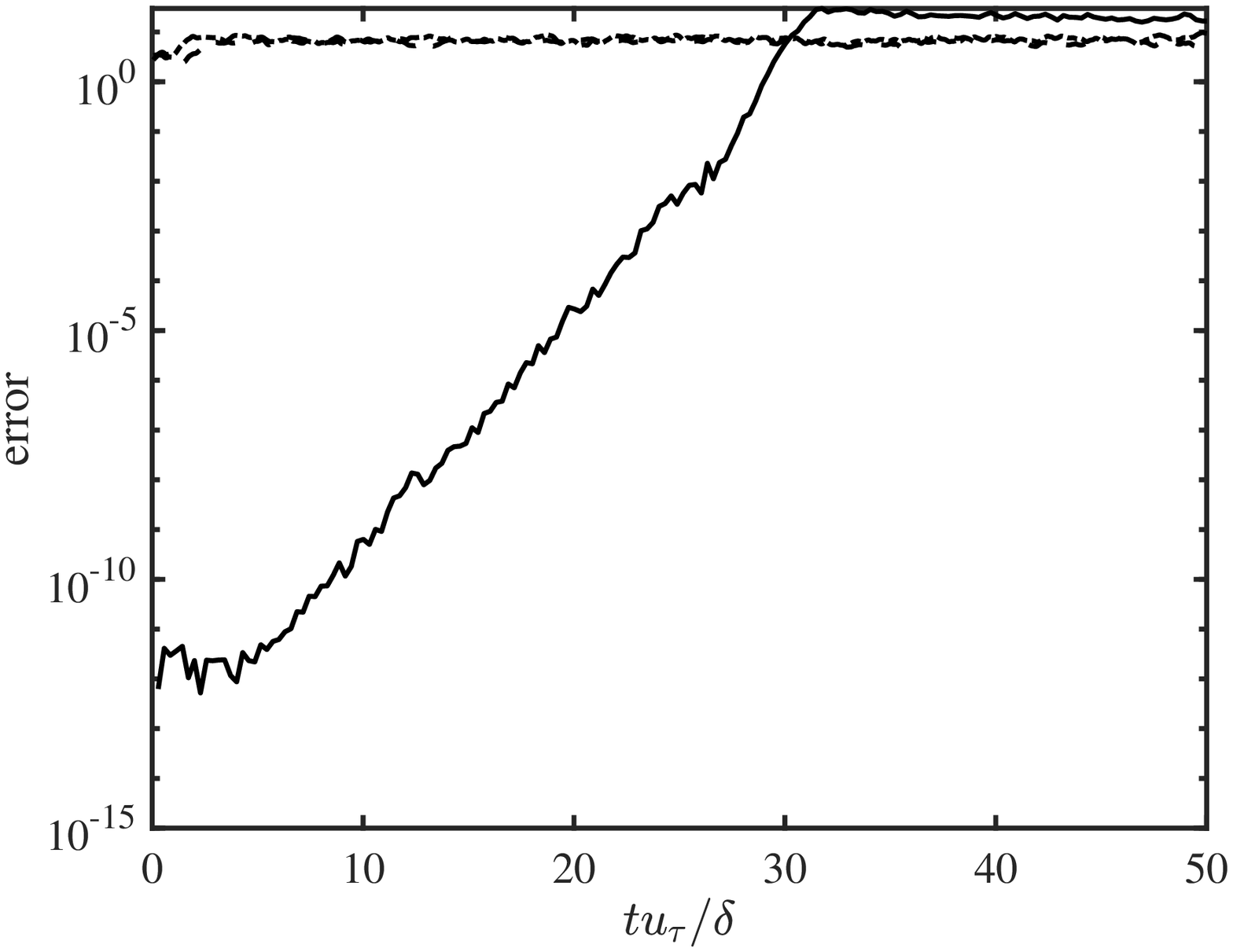}
\caption{
Error in (a) mean velocity profile and (b) streamwise turbulence
intensity  as a function of time for LES1 (\solidline), LES2
(\dashedline), and LES4 (\dotdashline) compared with DNS.
\label{fig:error}}
\end{center}
\end{figure}

The error in the mean velocity profile is computed as 
\begin{equation}
\text{error}(t) = \frac{\left[\int_0^\delta\left(\langle
u_1^{\mathrm{LES}}/u_\tau^{\mathrm{LES}}\rangle - \langle
u_1^{\mathrm{DNS}}/u_\tau^{\mathrm{DNS}}\rangle\right)^2\mathrm{d}x_2\right]^{1/2}}
{\left[\int_0^\delta\left(\langle
u_1^{\mathrm{DNS}}/u_\tau^{\mathrm{DNS}}\rangle\right)^2\mathrm{d}x_2\right]^{1/2}}, 
\end{equation}
where $\langle\cdot\rangle$ denotes averaging in homogeneous
directions. Figure \ref{fig:error}(a) shows the error for the three
LES cases as a function of time. For case LES1, the error in the mean
velocity profile grows exponentially from machine precision after
approximately $5\delta/u_\tau$ until it saturates at
$30\delta/u_\tau$. Note that the deviation from machine precision is
delayed compared with the instantaneous velocity error, indicating that
the requirement of SGS models to predict the correct
low-order statistics is less rigid than that for matching the
instantaneous velocity fields. Since SGS models are not expected to
retrieve exact instantaneous flow fields, the prediction of correct
low-order statistics would be sufficient for most industrial
applications.  Cases LES2 and LES4 show a second-order growth in
error that saturates at approximately $1\delta/u_\tau$.    

\begin{figure}
\begin{center}
\vspace{0.5cm}
\includegraphics[width=0.48\textwidth]{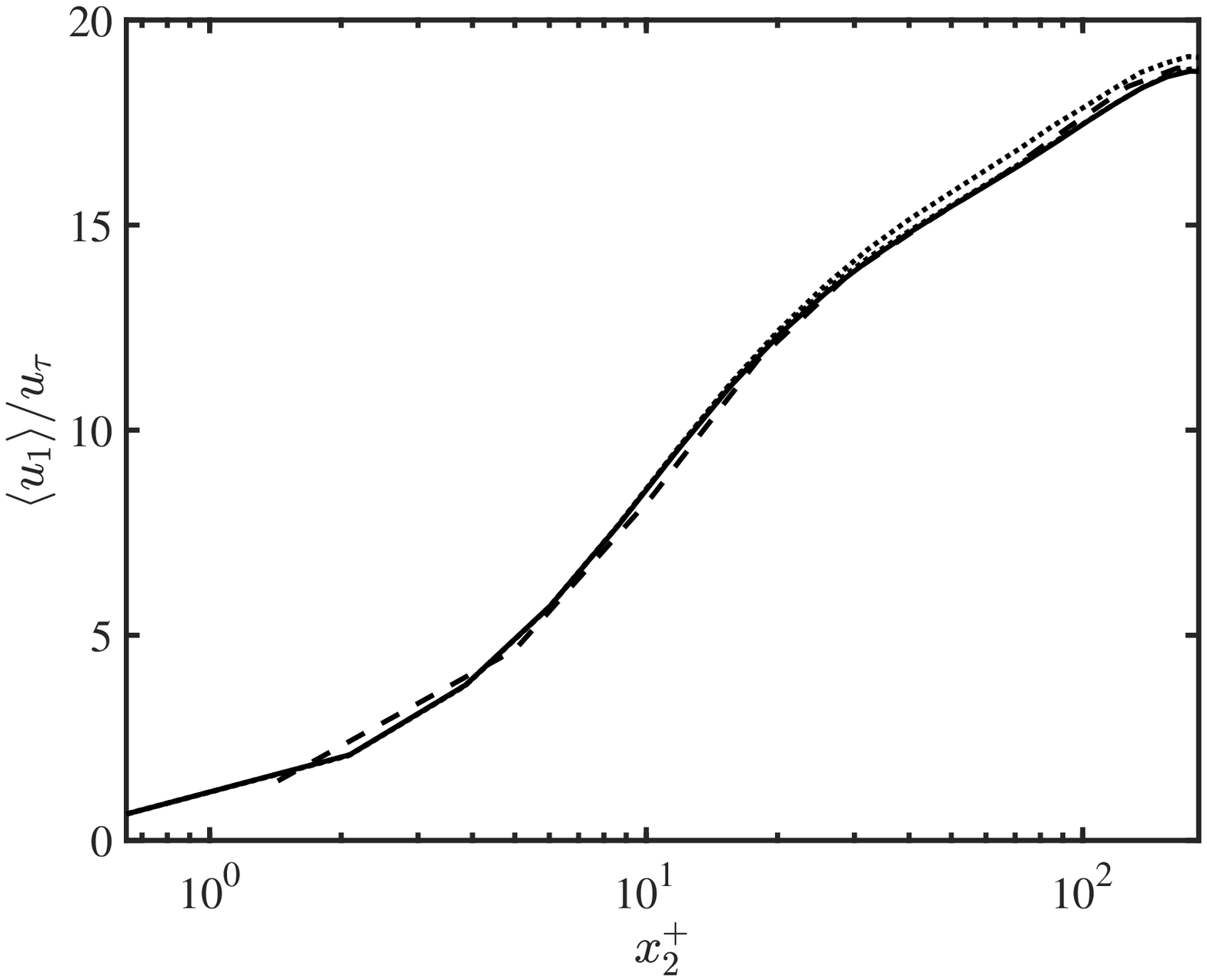}
\hspace{0.2cm}
\includegraphics[width=0.48\textwidth]{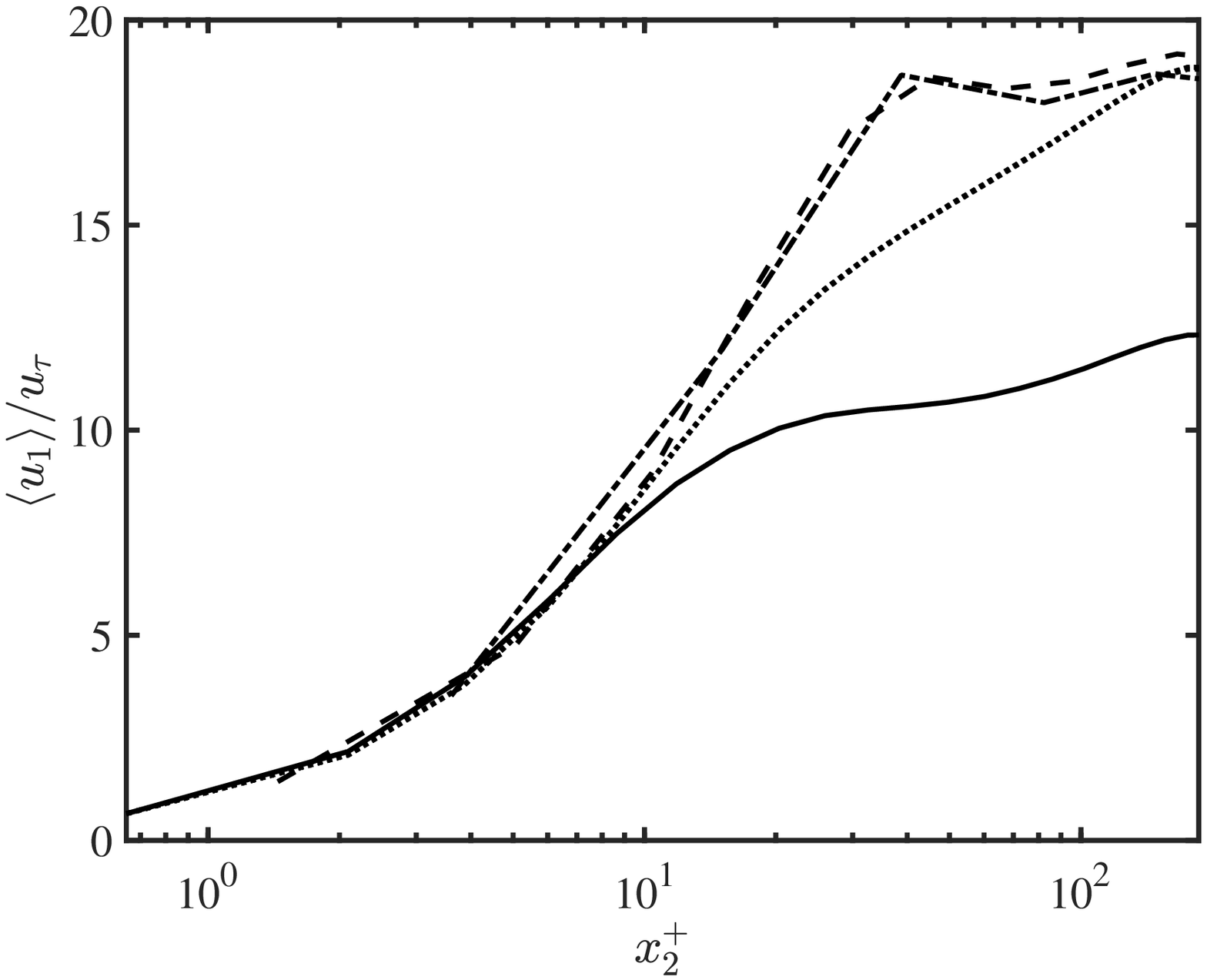}
\caption{
Mean velocity profile (a) before and (b) after saturation of error for
LES1 (\solidline), LES2 (\dashedline), and LES4 (\dotdashline)
compared with DNS (\dotline).
\label{fig:Umean}}
\end{center}
\end{figure}

Figure \ref{fig:Umean} shows the mean velocity profiles of the LES
cases before (for LES1, $tu_\tau/\delta < 20$; for LES2,
$tu_\tau/\delta < 1$) and after saturation of error. This shows that
while the corresponding DNS flow field can be useful in forming SGS
models, once the numerical errors compound and the DNS flow field
($u_i$ that form $\mathcal{T}_{ij}$) becomes uncorrelated to the LES
flow field ($\tilde{\bar{u}}_i$) at the corresponding time step, it
cannot provide information necessary to build effective SGS models for
flow statistics. 

%--------------------------------------------------------%
%--------------------------------------------------------%
%--------------------------------------------------------%
%--------------------------------------------------------%

\subsection{Turbulence intensities \label{sec:rms}}

\begin{figure}
\begin{center}
\vspace{0.5cm}
\includegraphics[width=0.48\textwidth]{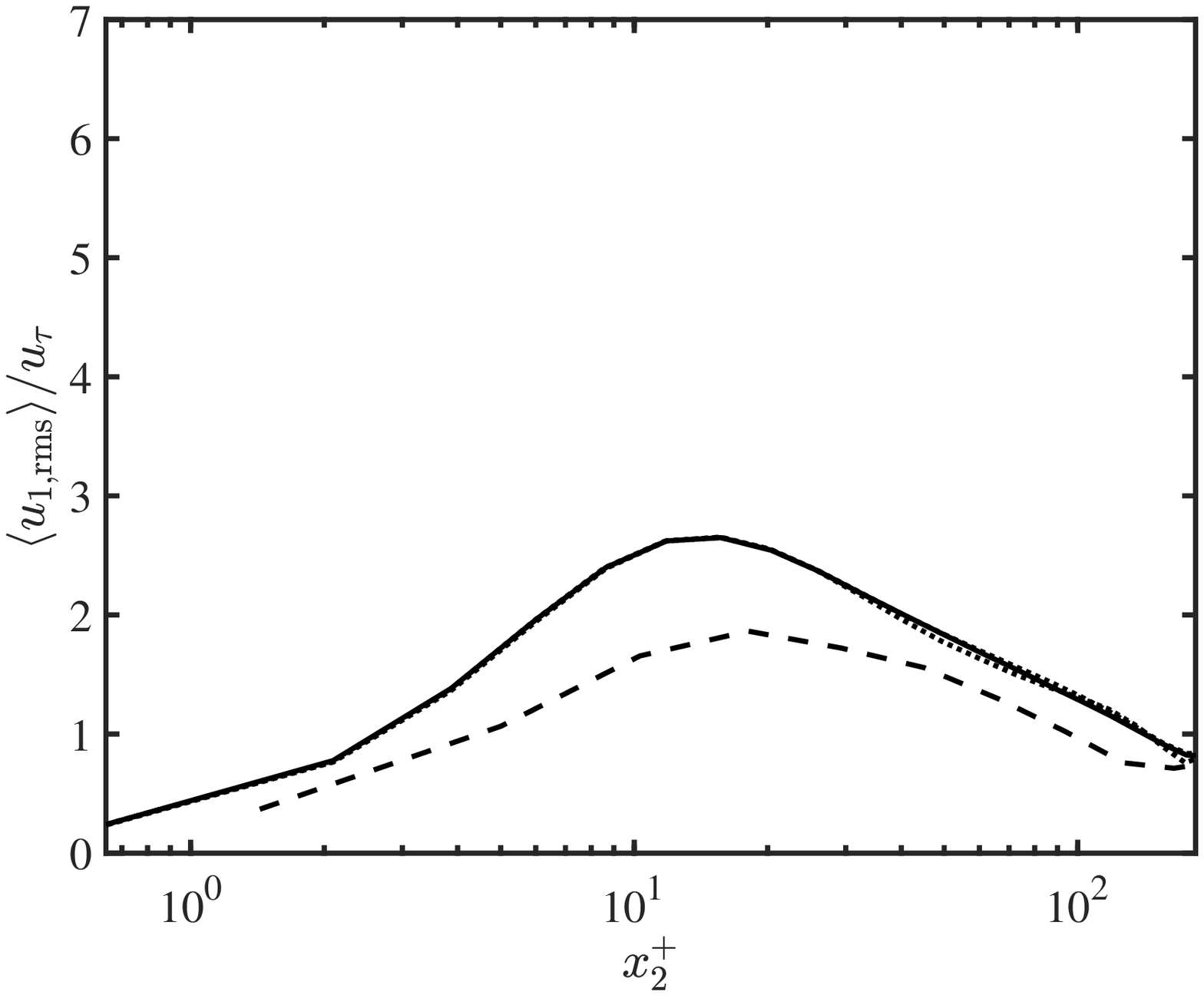}
\hspace{0.3cm}
\includegraphics[width=0.48\textwidth]{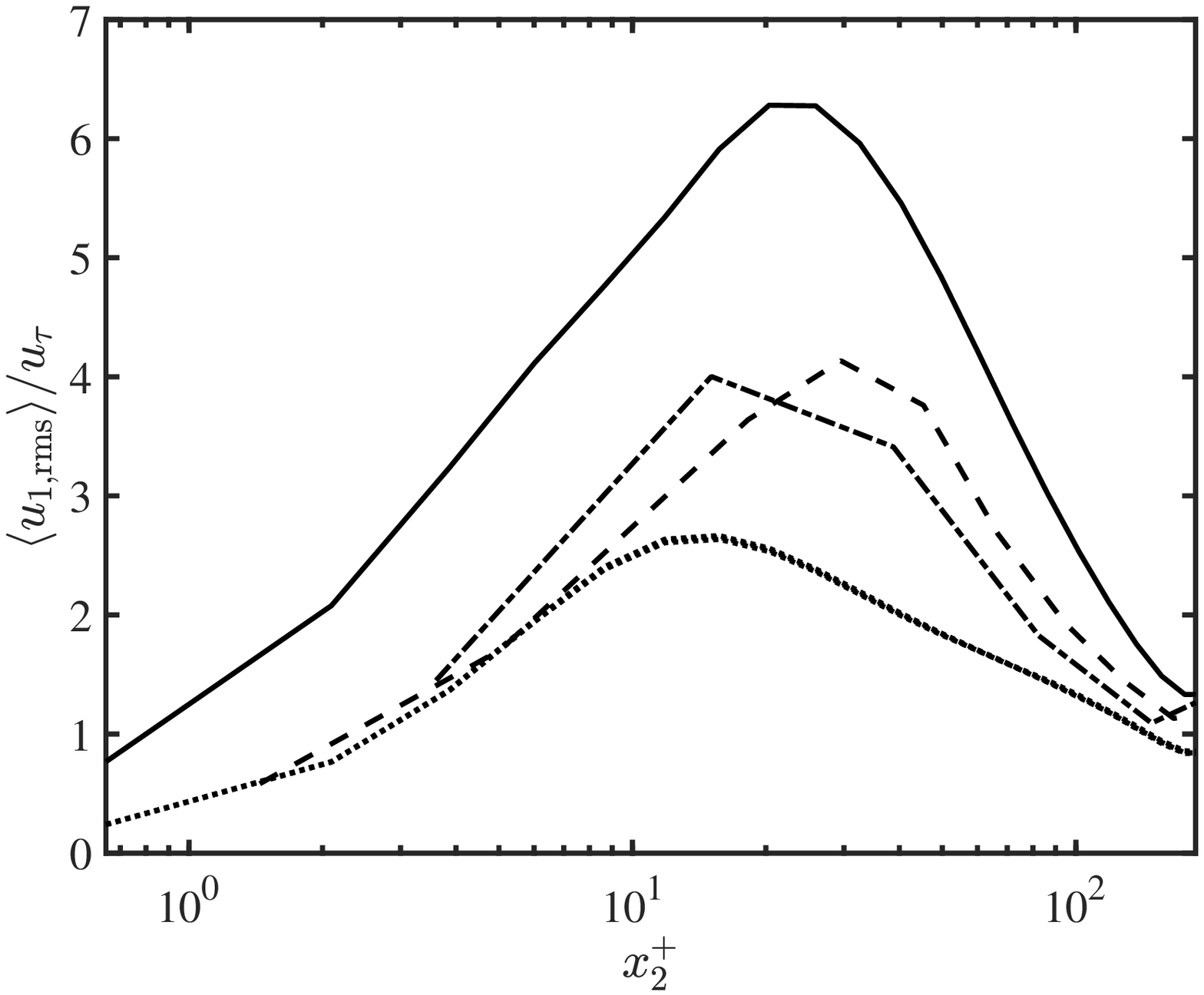}
\caption{
Streamwise turbulence intensity profile (a) before and (b) after
saturation of error for LES1 (\solidline), LES2 (\dashedline), and
LES4 (\dotdashline) compared with DNS (\dotline).
\label{fig:urms}}
\end{center}
\end{figure}

The error in the turbulence intensities is defined as 
\begin{equation}
\text{error}(t) = \frac{\left[\int_0^\delta\left(\langle
u_{i,\mathrm{rms}}^{\mathrm{LES}}/u_\tau^{\mathrm{LES}}\rangle - \langle
u_{i,\mathrm{rms}}^{\mathrm{DNS}}/u_\tau^{\mathrm{DNS}}\rangle\right)^2\mathrm{d}x_2\right]^{1/2}}
{\left[\int_0^\delta\left(\langle
u_{i,\mathrm{rms}}^{\mathrm{DNS}}/u_\tau^{\mathrm{DNS}}\rangle\right)^2\mathrm{d}x_2\right]^{1/2}}, 
\end{equation}
where rms denotes root-mean-squared quantities. The streamwise
turbulence intensity error is depicted in Figure \ref{fig:error}(b).
Although not shown, the wall-normal and spanwise turbulence intensity
errors show similar behavior. The errors in turbulence intensities are
comparable to that of the mean velocity profile except for an
immediate saturation of error for the LES2 and LES4 cases. This can be
explained by the missing scales in the coarser LES cases, as evident
from the streamwise turbulence intensity profile before saturation of
error in Figure \ref{fig:urms}(a). The mismatch between the DNS
profile and the profile of case LES2 before the saturation of error
can be explained by the energy contained in the small scales that are
present in the DNS resolution but missing in the LES2 resolution,
which is expected from true LES. A fair comparison can be made by
discretizing the DNS flow field to the LES resolution, effectively
removing the small scales, or by taking into account the SGS
contribution for the LES case. Similar to the mean velocity profile,
after the saturation of error, the DNS flow field no longer is able to
provide the correct SGS contribution for the LES; thus the
turbulence intensity predictions diverge from the true statistics. 

%--------------------------------------------------------%
%--------------------------------------------------------%

\section{Summary \label{sec:summary}}

The equations for LES are formally derived by low-pass filtering the
NS equations with the effect of the small scales on the larger ones
captured by a SGS model. However, it is known that the LES equations
usually employed in practical applications are inconsistent with the
filter operator when no explicit filter is used. Moreover, even for
explicitly filtered LES, some inconsistencies remain in the
wall-normal direction owing to the constraining effect of the wall. A
typically undesirable effect stemming from the inconsistency between
LES and the fNS equations is that DNS data cannot be used to aid SGS
modeling.

Using the proposed form of the incompressible fNS equations for flows
over flat walls in which the continuity equation, the filter operator,
and the boundary condition at the wall are consistently formulated, we
compute a set of plane turbulent channel flow explicitly filtered LES
at different grid resolutions with the SGS term computed from a
concurrent DNS computation starting with an equivalent initial
condition. The results show that the SGS term computed from DNS
provides a truthful SGS model for short times until the numerical
errors compound and the two parallel computations diverge. The
effective time when the statistics remain similar is longer for
lower-order statistics such as mean velocity profile, as expected.
This demonstrates that with the consistent fNS equations, the abundant
DNS data can be used to inform SGS model development for explicitly
filtered LES through methods such as machine learning. Moreover, the
diverging of statistics at longer times shows that the SGS
contribution computed from an uncorrelated DNS flow field is
ineffective. This shows that the SGS contribution is not necessarily
homogeneous as most SGS models assume, and that a more precise correlation
between the instantaneous LES velocity field and the SGS model is required
to capture the effect of the small scales on the resolved flow. 

% Future work
%Commuting discretization and differentiation operator
%Extension method for wall curvature

\section*{Acknowledgments} 

This work was supported by NASA under the Transformative Aeronautics
Concepts Program, grant no. NNX15AU93A.

%=====================================================================
%=====================================================================
\bibliographystyle{ctr}
%\bibliography{eLES.bib}

%\bibliographystyle{ctr}
%\bibliography{eLES.bib}
%\begin{thebibliography}{} 
%\end{thebibliography} 

%%%%%%%%%%%%%%%%%%%%%%%%%%%%%%%%%%%%%%%%%%%%%
\end{document}